\documentclass[journal]{IEEEtai}

\usepackage{amsmath,amsfonts}
\usepackage{dsfont}
\usepackage{algorithm}
\usepackage{setspace}
\usepackage[noend]{algpseudocode}

\usepackage[commentColor=black]{algpseudocodex}
\tikzset{algpxIndentLine/.style={draw=black}}
\algrenewcommand{\alglinenumber}[1]{\bfseries\footnotesize #1}
\algrenewcommand{\algorithmicrequire}{\textbf{Input:}}
\algrenewcommand{\algorithmicensure}{\textbf{Output:}}

\let\labelindent\relax

\usepackage{array}
\usepackage{multirow, array, ragged2e, lscape, eqparbox, xspace, setspace}
\usepackage{booktabs, tabularx,tabulary}

\usepackage[caption=false,font=normalsize,labelfont=sf,textfont=sf]{subfig}
\usepackage{graphicx}

\usepackage[colorlinks = true,
            linkcolor = blue,
            urlcolor  = blue,
            citecolor = blue,
            anchorcolor = blue]{hyperref}
\usepackage{url}
\usepackage{cite}
\usepackage{orcidlink}

\usepackage{textcomp}
\usepackage{stfloats}
\usepackage{verbatim}
\usepackage{makecell}
\usepackage{pifont}
\usepackage{enumitem}  
\usepackage{textcomp}

\graphicspath{{figures/}}

\usepackage{hyphenat}

\renewcommand{\vec}[1]{\mathbf{#1}}

\makeatletter
\DeclareRobustCommand\onedot{\futurelet\@let@token\@onedot}
\def\@onedot{\ifx\@let@token.\else.\null\fi\xspace}

\def\etal{\emph{et al}\onedot}

\makeatother

\begin{document}

\title{Explainability-Based Adversarial Attack on Graphs Through Edge Perturbation}

\author{Dibaloke Chanda \orcidlink{0000-0001-5993-659X},~\IEEEmembership{Student Member,~IEEE}, \\ 
Saba Heidari Gheshlaghi \orcidlink{0000-0002-8252-3686},~\IEEEmembership{Student Member,~IEEE},\\
Nasim Yahya Soltani \orcidlink{0000-0002-4502-8715},~\IEEEmembership{Member,~IEEE}
\thanks{}}



\maketitle

\begin{abstract} Despite the success of graph neural networks (GNNs) in various domains, they exhibit susceptibility to adversarial attacks. Understanding these vulnerabilities is crucial for developing robust and secure applications. 
In this paper, we investigate the impact of test time adversarial attacks through edge perturbations which involve both edge insertions and deletions. A novel explainability-based method is proposed to identify important nodes in the graph and perform edge perturbation between these nodes. The proposed method is tested for node classification with three different architectures and datasets. The results suggest that introducing edges between nodes of different classes has higher impact as compared to removing edges among nodes within the same class.

\end{abstract}


\begin{IEEEImpStatement}
The task of node classification in GNNs has a substantial effect on tasks that involve network analysis in numerous domains. Considering the broad applicability of this method, understanding potential strategies for adversarial attacks can provide insight to defend against them. In this work, we propose a novel attack on GNNs based on explainability. Explainability offers comprehensive reasoning behind the predictions made by GNNs and facilitates transparency about the inner operation of the model. We show that additional information and insights that can be gained through GNN-based explainability methods can be utilized to strengthen the adversarial attack. Performing adversarial attacks on an important subgraph determined by the explainability algorithm is more influential than adversarial attack on the entire graph. In addition, adversaries can strategically change the graph structure by having more edge insertions compared to edge deletion to increase the impact of the attack.


\end{IEEEImpStatement}

\section{Introduction}\label{sec:Introduction}
{\IEEEPARstart{G}NNs have recently received significant attention for their impressive performance across different fields, such as social network analysis, biology, chemistry, recommendation systems, and healthcare. These deep learning (DL) models are designed to handle complex graph-structured data on various tasks such as node classification, connection prediction and graph classification \cite{wu2020comprehensive,zhou2020graph}. 

Even though DL models have achieved remarkable success in various domains, recent studies have shown their vulnerability to adversarial attacks. These attacks can drastically undermine the models' effectiveness and performance by introducing small modifications to the input data, alterations that are typically imperceptible to humans, yet can cause the model to make wrong predictions. \cite{attack2, xu2020adversarial}. Adversarial attacks can reveal significant weaknesses in the model's generalization ability and raise concerns about the reliability of its decision-making in real-world applications. As such, understanding and mitigating adversarial attacks is crucial for maintaining the integrity and robustness of DL models, including those that handle complex inputs like images, sound, and graph-structured data. Similarly, GNNs, as part of the DL model family, are not immune to such adversarial attacks. In GNN, the attacker can create the adversarial perturbation by manipulating and changing the graph structure, connections and node features \cite{attack1, xu2019topology,zang2020graph,tao2020adversarial, ma2019attacking}. 

As GNNs advance into application areas where security and reliability are of paramount importance, the need for explainable models that can withstand adversarial conditions is imperative. Explainability in GNNs refers to the methods designed to uncover the reasoning behind the predictions made by networks. This involves tracing the flow of information across the graph structure and understanding how individual node relationships and graph substructures contribute to the model's performance and output. The pursuit of explainability not only aids in validating the model's decisions but also ensures transparency, reliability, and identifies key nodes and edges that may be more vulnerable to attack \cite{yuan2022explainability, agarwal2023evaluating, xu2021explainability}. 

As highlighted, for adversarial attacks to remain undetected by humans, it is crucial that the number of modifications made during the attack is minimized, thereby ensuring the changes are as inconspicuous as possible. To ensure the attack is unnoticeable, the number of possible modifications, such as adding or deleting edges, is restricted by a budget~\cite{advg1}. Yet, it is important to note that due to the graph's properties, even within these constraints, such actions can still result in changes to the graph's structure that may be noticeable. 

\section{Literature Review} \label{sec:Literature_Review}

Over the past few years, research studies have generated a substantial volume of literature dedicated to the exploration of adversarial attacks on GNNs. The research indicates a growing interest in understanding and mitigating vulnerabilities within these state-of-the-arts models, which are increasingly employed for tasks that interpret graph-structured data. Studies have shown how adversarial attack can manipulate the predictive accuracy of GNNs by introducing subtle, often imperceptible perturbations to the graph data. As these DL models become integral to more applications, the significance of study on methods for generating adversarial samples escalated~\cite{chen2020survey,jin2021adversarial, sun2022adversarial, xu2018characterizing}. Jin et al. \cite{advg6} demonstrated that even small perturbation could lead to a significant decrease in test accuracy. This research investigated various attack strategies, revealing that the adding just a few edges to the input graph could significantly impact the GNN models' performance.
Z{\"u}gner et al. \cite{advg1} pioneered the exploration of adversarial vulnerabilities in GNN, proposed an adversarial attack approach called Nettack, employing a surrogate model to generate adversarial samples. This approach iteratively generates adversarial samples, modifying them based on changes in confidence values resulting from added perturbations. While this technique stands as a forerunner in adversarial attacks on GNNs, its applicability to large-scale graphs is limited due to the high time complexity. In the study by Dai et al. \cite{dai2018adversarial}, a reinforcement learning approach was employed to apply a black-box node-level attack. Conversely, Entezari et al. \cite{entezari2020all} employed a first-gradient optimization technique for attacking GNN, coupled with low-rank approximation strategies to reduce the attack's impact and enhance the robustness of the network. Additionally, Chen et al. \cite{chen2018fast} introduced another gradient-based approach named projected gradient descent (PGD), where gradients of edges are extracted and the highest absolute gradients between pair of nodes used to generate the attack strategy. In the work of Zang et al. \cite{zang2020graph}, the researchers focused on targeted attacks, postulating the existence of ``bad actor" nodes. This paper demonstrated that fliping the edges between these nodes and any the target node, the network performance on targeted node will be negatively affected.
Sun et al. \cite{sun2020adversarial} employed a surrogate model to tackle the bi-level optimization problem, applying meta-gradient techniques for the addition or deletion of edges. They additionally showed a slight tendency for the algorithm to link nodes with lower degrees.

None of the research work other than~\cite{xu2021explainability} explores the use of GNN specific explainability methods to gain additional information about the graph structure and utilize it to amplify the effectiveness of adversarial attacks. In addition, most of the work treats all edges with similar importance rather than giving a specific subset of edges more importance. 
The contributions of this  work can be summarized as:
\begin{enumerate}
    \item  A novel explainability-based approach is proposed to identify important nodes and form an important subgraph from the original input graph.
    
    \item The effects of edge perturbations among nodes of the important subgraph is evaluated and used to propose a novel adversarial attack.

\end{enumerate}

The remaining parts of the paper are organized as follows. Section~\ref{sec:background}, provides a brief description of GNN architectures in addition to a background on the adversarial attack complexity and categorization in GNN. Section~\ref{sec:methodology}, outlines the mathematical framework training, inference, and rewiring strategy, along with details about the capabilities and knowledge of the attacker. In section~\ref{sec:num_test}, dataset description, hyperparameters of the model and the numerical test results are presented followed by section~\ref{sec:conclusion} which provides a conclusion.



\section{Background on GNN and Adversarial Attacks on GNN}\label{sec:background}
In this section, we introduce GNNs architectures used in this work and go over categorizations of adversarial attacks on GNN. 

\subsection{Architectures of GNNs}
A GNN is constructed of $L$ permutation equivariant layers where $l \in \{1,2, \cdots, L\}$ refers to $l^{th}$ layer. The message-passing algorithm in a GNN can be represented by \eqref{message_passing}. A node $u$ aggregates information from its neighboring nodes $v \in \mathcal{N}(u)$ and updates its current state where $\mathcal{N}(u)$ represents the set of neighboring nodes for node $u$. This entire message-passing scheme happens in a three-step process. First, the embeddings of the target node and neighboring nodes of the current layer go through an affine transformation which can be denoted by a differentiable function $\phi^{(l)}$.

\begin{equation}
\mathbf{h}_u^{(l)}=\gamma^{(l)}\left(\mathbf{h}_u^{(l-1)}, \bigoplus_{v \in \mathcal{N}(u)} \phi^{(l)}\left(\mathbf{h}_u^{(l-1)}, \mathbf{h}_v^{(l-1)}\right)\right)
\label{message_passing}
\end{equation}
Then, an aggregation function aggregates the hidden state of the neighboring nodes which can be done via a permutation invariant differentiable function $\bigoplus$. Finally, the aggregated hidden state is passed through an update function $\gamma^{(l)}$ to get the updated hidden representation for target node $u$ for the current $l^{th}$ layer. 
 Based on this scheme there are several GNN variants based on the choice of aggregation function, permutation invariant differentiable function and the update function.

\subsubsection{Graph Convolutional Network (GCN)}
GCN is a special variant of the GNN first introduced by~\cite{kipf2016semi}.
\begin{equation}
\label{eq:gcn_eq}
\mathbf{h}_u^{(l)}=\sigma\left(W^{(l)} \sum_{v \in \mathcal{N}(u) \cup\{u\}} \frac{\mathbf{h}^{(l-1)}_v}{\sqrt{\lvert \mathcal{N}(u)||\mathcal{N}(v)\rvert}}\right)
\end{equation}

The message-passing scheme for this specific method is given by \eqref{eq:gcn_eq}. Here, $\frac{1}{\sqrt{\lvert \mathcal{N}(u) \mid \mid \mathcal{N}(v) \rvert }}$ is the normalization factor, $\mathbf{h}_{v}$ is the hidden representation of the neighboring nodes, $W^{(l)}$ is the weight matrix consisting trainable parameters for the $l^{th}$ layer, $\sigma$ is the non-linear activation function. Note that, $\mathcal{N}(\cdot)$ is used to denote the neighboring node the $\sum$ operator is chosen as the permutation invariant aggregation function. After the message passing stage,  the hidden representation of the target node $\mathbf{h}^{(l)}_{u}$ is generated for the current layer $l$.

\subsubsection{Graph Attention Network (GAT) }

GAT~\cite{velivckovic2017graph} make use of the attention mechanism introduced by Vaswani~\etal \cite{vaswani2017attention} to generate attention score between nodes in the graph. To get the attention score $\alpha_{ij}$ where $i$ is the index of the target node and $j$ is the index of one of the neighboring nodes \eqref{eq:GAT} is utilized.

\begin{equation}
\label{eq:GAT}
\alpha^{(l-1)}_{i j}=\frac{\exp \left(\sigma\left(\mathbf{a}^{{(l-1)}^{\top}} \left[W \vec{h}_i^{(l-1)} \| W \vec{h}_j^{(l-1)}\right]\right)\right)}{\sum_{k \in \mathcal{N}_i} \exp \left(\sigma\left(\mathbf{a}^{{(l-1)}^{\top}} \left[W\vec{h}^{(l-1)}_i \| W\vec{h}_k^{(l-1)}\right]\right)\right)}
\end{equation}
Where $\mathbf{h}_i$ is the hidden representation of the target node, $\mathbf{h}_j$ is the hidden representation for one of the neighboring nodes, $\mathbf{a}^{\top}$ is the shared learnable weight vector. After applying the non-linear activation $\sigma$, the unnormalized attention score between node $i$ and node $j$ is generated. To normalize the score it is passed through a softmax layer to get the normalized attention score $\alpha_{ij}$. Note that here ${.}^{\top}$ is the transposition operation and $\|$ is concatenation.

\begin{equation}
\label{eq:GAT2}
\vec{h}_i^{(l)}=\sigma\left(\sum_{j \in \mathcal{N}_i} \alpha^{(l-1)}_{i j} W^{(l)} \vec{h}^{(l-1)}_j\right)
\end{equation}

Finally, the attention score is used to compute the hidden representation for node $i$ according to \eqref{eq:GAT2}.

\subsubsection{GraphSAGE} 

GraphSAGE which stands for graph sample and aggregate was introduced by \cite{hamilton2017inductive} and is a popular method for training large graphs. Even though accuracy-wise it is inferior to GAT and GCN, it has the advantage of faster training time. The message passing scheme for GraphSAGE is given by \eqref{eq:sage}.

\begin{equation}
\label{eq:sage}
\mathbf{h}^{(l)}_{u}= \sigma \left( W_{1}^{(l-1)} \mathbf{h}^{(l-1)}_{u}+W^{(l-1)}_{2} \frac{1}{|\mathcal{N}(u)|} \sum_{v \in \mathcal{N}(u)} \mathbf{h}^{(l-1)}_{v} \right)
\end{equation}

where, $W^{(l-1)}_{1}$ and $W^{(l-1)}_{2}$ are two different weight matrices, $\mathbf{h}^{(l-1)}_{u}$ is the hidden representation of the target node from the previous layer and $ \mathbf{h}^{(l-1)}_{v}$ is the hidden representation of the neighboring nodes from the previous layer.

 \subsection{Adversarial Attack on GNN}\label{subsec:background_at}
 Conducting adversarial attacks on GNNs presents a higher level of complexity compared to other DL models, and this can be attributed to multiple challenges in the graph domain such as :
\begin{enumerate}
\item {\textbf{Complex perturbation space:} unlike perturbations in standard data types, perturbations in the graph domain are inherently more complicated. Attackers have different ways to apply the attack and can modify the nodes and edges such as changing node features, inserting or deleting nodes, as well as adding or removing edges to generate adversarial samples. This multiplicity of attacks is due to the flexible nature of graph data, where the structural and feature-related aspects of the data can be affected in various ways to attack the model \cite{wang2018attack}.}

\item {\textbf{Discrete data domain:} graph data is discrete by nature, which stands in contrast to the continuous data often processed by other DL frameworks. This discreteness adds an additional layer of difficulty to the optimization processes involved in crafting effective adversarial examples, as gradient-based techniques, which are widely used for continuous data, may not be directly applicable or as effective in the discrete terrain.}
\begin{figure*}[!t]
    \centering
    \includegraphics[width=0.79\linewidth]{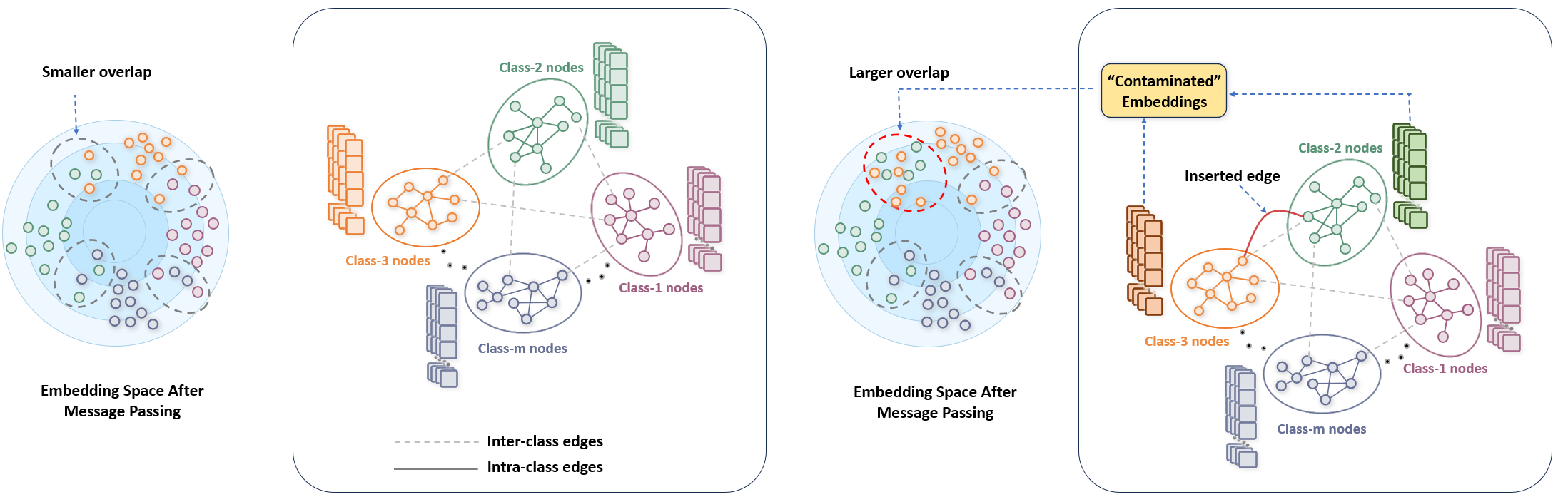}
    \caption{Visualization of edge insertion between nodes of different classes contaminating the node embeddings rather than refinement. (\textbf{Left}) Small overlap between embedding from different classes and easily separable. (\textbf{Right}) ``Contaminated Embeddings" of class-2 and class-3 after intra-class edge insertion; larger overlap and no longer easily separable.}
    \label{fig:idea}
\end{figure*}

\item {\textbf{Interdependent data characteristics:} GNNs are designed to take the relational dependencies inherent in graph-structured data into account. Each node is understood in context, with its features and connections informing its representation in relation to other nodes. This interconnectivity means that any perturbation does not simply affect a single instance, but rather has a cascading impact on the model’s interpretation of interconnected nodes. As a result, the adversarial strategies must consider these dependencies, further complicating the attack process.}
\end{enumerate}

Hence, researchers are prompted to develop more advanced techniques to design adversarial attacks and robustness on GNNs. In addition, adversarial attacks 
can have different levels of knowledge and access as summarized below.

\begin{enumerate}

\item{\textbf{White-box, gray-box and black-box attacks:} white-box attacks are the type of attack when the attacker has complete knowledge and understanding of the neural network's architecture. In a white-box attack, the attacker has access to all the details, including the model's parameters, training data, and their corresponding labels. In a white-box attack, the attacker generates adversarial samples that are precisely tuned to the model's vulnerabilities, often with a high degree of success. In the gray-box attack, the attacker has access to limited information about the victim's model, making this attack potentially more dangerous than a white-box attack. For instance, the attacker can access the model's parameters but not its training data. On the other hand, black-box attacks are the type of attacks where the attacker has restricted knowledge. In the black-box attack, the internal workings of the model, including its parameters and training data are not visible and accessible to the attacker. In this type of attack, the attacker may only have access to the model's inputs and outputs, and generate the attack by only relying on this limited information \cite{wu2019adversarial}.}
\item{\textbf{Targeted vs. untargeted attacks:} another common adversarial attack categorization is to distinguish between targeted and untargeted approaches. Targeted attacks are when the adversary's goal is to manipulate the model in order to misclassify inputs into specific, incorrect categories. As regards untargeted attacks, they are less discriminative and aim to reduce the network's overall accuracy and performance, leading to a general state of unreliability in the model's predictions.}
\item{\textbf{Evasion vs. poisoning attacks:} another important aspect in adversarial attack is timing and when the attack is applied to the network. Evasion attacks occur after the model is trained, where the attacker introduces adversarial inputs designed to be misclassified by the already trained model. In contrast, poisoning attacks happen during the training or even before the model's training procedure. In this case, the attacker applies adversarial attack in the training phase; hence, the learning process is corrupted and the network fails to perform accurately even before training \cite{tang2020transferring, liu2019unified, zhou2019data, zhang2019towards}.}
\end{enumerate}
Based on the categorization outline above, our proposed method is a  white-box, untargeted and evasion type adversarial attack. The rationale and the justification behind deciding each category are provided in section~\ref{prev} in detail.
\begin{figure}[!h]
    \centering
    \includegraphics[width=0.96\linewidth]{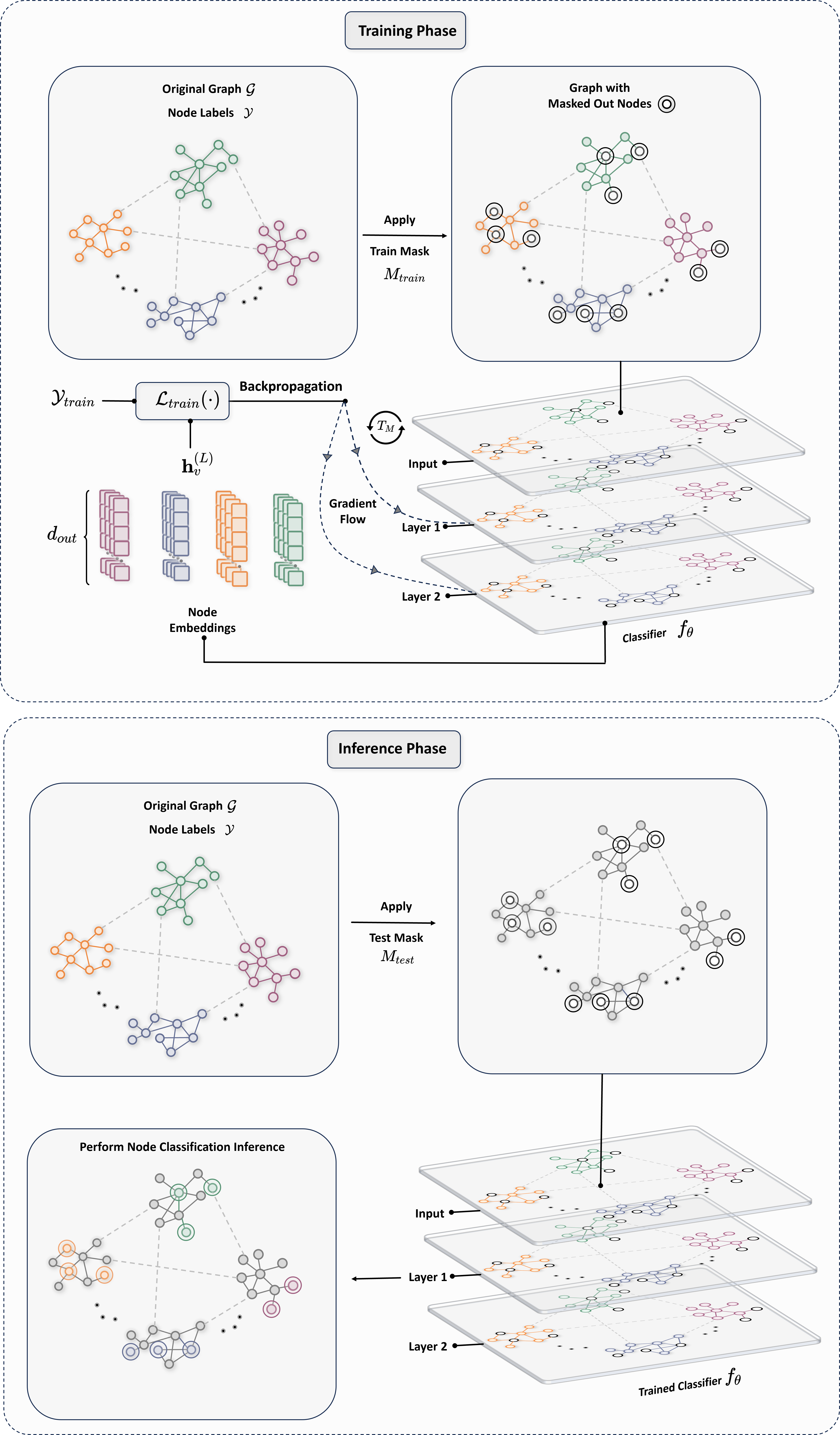}
    \caption{GNN training and inference for node classification under transductive settings. (\textbf{Top})~the node labels for test nodes are masked out with a train mask. The masked-out graph is passed through the GNN layers for training. Loss is calculated with the help of ground truth labels and model parameters are updated with backpropagation. (\textbf{Botttom}) In the inference stage, the test mask is applied followed by a forward pass through the trained model and inference for node classification is performed for the test node labels.}
    \label{fig:node_classification_training}
\end{figure}

\section{Methodology}\label{sec:methodology}

Our proposed algorithm builds upon the idea that is visualized in Fig.~\ref{fig:idea}. We argue that when the message-passing algorithm is carried out inside the GNN layers, features that propagate between nodes of the same class play a crucial role in refining the node embeddings. 
Edge deletion between the nodes of the same class will halt this process. But more importantly, introducing edges between nodes of different classes will ``pollute" or ``contaminate" the features of different classes when the message-passing takes place and hence generating noisy embeddings which is not representative of the nodes of that class. Without any edge perturbations, the embeddings generated after the message-passing scheme are nicely separable with minimal overlap. But with edge perturbations, especially with edge insertions, the generated embeddings are no longer easily separable in the embedding space. The goal of the adversary is to increase this overlap as much as possible with unnoticeable modification to the graph structure. 
 
\subsection{Mathematical Framework for Training and Inference}
The input graph is defined as $\mathcal{G}(\mathcal{V},\mathcal{E},\mathbf{X})$ where $\mathcal{V}$ is the node set, $\mathcal{E} \subseteq \mathcal{V}\times \mathcal{V}$ is the edge set and $\mathbf{X} \in \mathbb{R}^{|\mathcal{V}| \times d_{in}}$ is the feature matrix. The total number of nodes is $|\mathcal{V}|=N$ and the dimension of each node feature vector is $d_{in}$. There is an associated label for each node $v \in \mathcal{V}$ which leads to the label set $\mathcal{Y}$. The label set $\mathcal{Y}$ can broken down into $m$ non-overlapping sets where $m$ is the number of classes. Mathematically, that can be expressed as the following equation.

\begin{equation}
    \mathcal{Y}:=\bigcup^{m}_{i=1} \mathcal{Y}_{i} 
\end{equation}
Each $\mathcal{Y}_i$ contains the node label for nodes in class $i$. We consider three different classifiers which can be expressed in a set as $\{f_{\theta}^{GCN},f_{\theta}^{GAT}, f_{\theta}^{GraphSAGE}\}$ where $f_{\theta} \in \{f_{\theta}^{GCN},f_{\theta}^{GAT}, f_{\theta}^{GraphSAGE}\}$ is used to refer to one of the classifiers parameterized by $\theta$.  The overall process of training and inference under transductive settings is presented in Fig.~\ref{fig:node_classification_training}. 

\begin{figure*}[!t]
    \centering
    \includegraphics[width=0.9\linewidth]{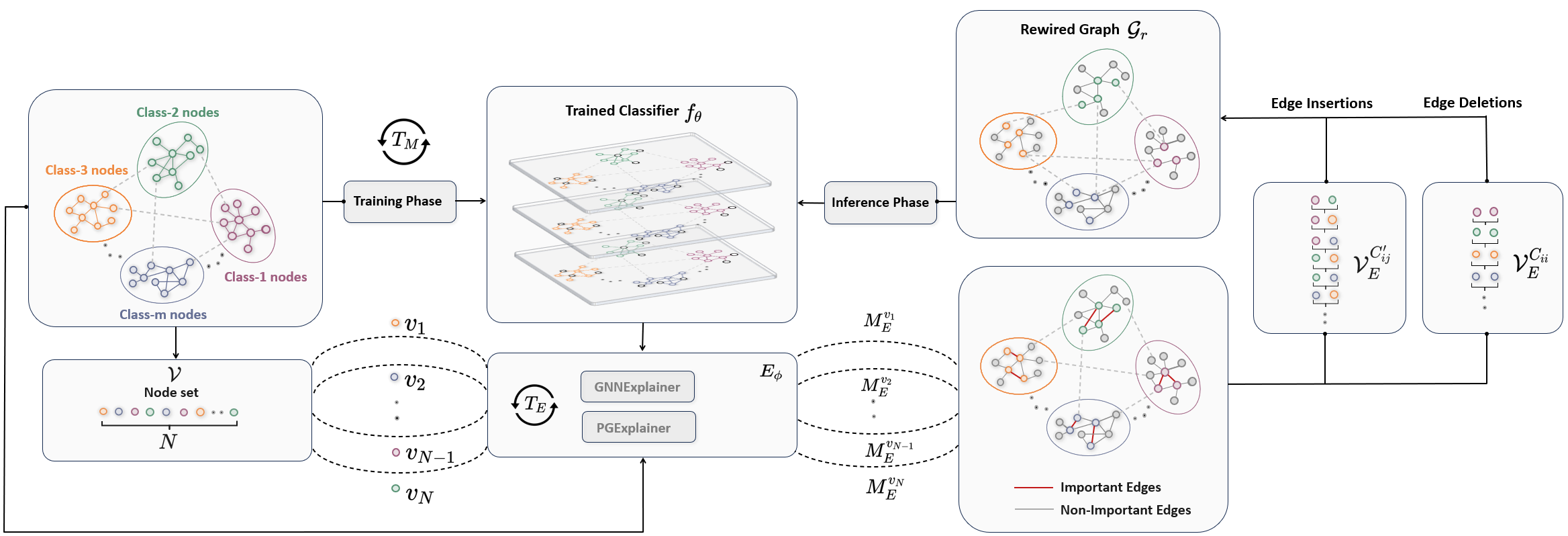}
    \caption{Proposed method to identify the important edges and important nodes in the graph with explainability and rewire the graph. Start with the input graph $\mathcal{G}$ and train it for $T_{M}$ epochs under transductive settings. Then for each node $v$ from the node set $\mathcal{V}$ run the explainability model $E_{\phi}$ optimization for $T_E$ epochs to generate explainability mask $M^{v}_{E}$. Finally, combine the generated explainability mask to get a combined mask $M_{E}$ and perform edge insertions and deletions to get the rewired graph $\mathcal{G}_r$.}
    \label{fig:explanainbility_rewire}
\end{figure*}

As this is a node classification task, there is an associated train mask $M_{train}\in \mathbb{R}^{|\mathcal{V}|}$ which is used to create the training node labels $\mathcal{Y}_{train}$. The graph  $\mathcal{G}$  is fed to the classifier $f_{\theta}$ and node embeddings are generated for each node $v\in \mathcal{V}$. The generated node embeddings can be expressed as $\left\{\mathbf{h}_v^{(L)} \in \mathbb{R}^{d_{out}}, \forall v \in \mathcal{V} \right\}$ where $L$ represents the last layer of the classifier $f_{\theta}$ and $d_{out}$ is the dimension of the generated node embeddings. The node embedding generation 
process can be expressed with  the following equation:

\begin{equation}
    \mathbf{h}_v^{(L)} := f_{\theta}(\mathcal{G})
\end{equation}

Note that, during the training stage, a validation mask $M_{val}\in \mathbb{R}^{|\mathcal{V}|}$ is used to generate $\mathcal{Y}_{val}$ (validation node labels) to assess the performance of the training with. The model with the lowest validation loss $\mathcal{L}_{val}$ is the model used in the inference stage. To backpropagate the errors in prediction during the training stage cross-entropy loss function $\mathcal{L}_{train}$ is used to compute the loss by taking in the generated embeddings $\mathbf{h}_v^{(L)}$ and training node labels $\mathcal{Y}_{train}$ as input.

In the inference stage, the test mask $M_{test} \in \mathbb{R}^{|\mathcal{V}|}$ is used to get the test nodes which have corresponding ground truth labels $\mathcal{Y}_{test}$. After applying the test mask node classification on the test nodes is performed.

\subsection{Mathematical Framework for Graph Rewiring}
The overall workflow of our proposed model is visualized in~Fig.~\ref{fig:explanainbility_rewire} and the algorithm is shown in Algorithm~\ref{algo:training}. Our graph rewiring method works by utilizing two explainability methods, GNNExplainer\cite{ying2019gnnexplainer} and PGExplainer\cite{luo2020parameterized}. 

GNNExplainer uses an optimization approach focused on increasing the mutual information between predictions and the potential subgraph structures' distribution. To be able to perform optimization in discrete graph structure they assume continuous relaxation of the graph adjacency matrix. Given an input graph and the model used to train for node classification, GNNExplainer is able to identify the important edges that contributed to the prediction task. The output of the explanability method is a continuous mask for all the edges. PGExplainer uses a similar approach to GNNExplainer. PGExplainer makes use of deep neural networks to parameterize the generation process of explanations where it is claimed that the algorithm has a better generalization capability under inductive settings~\cite{luo2020parameterized}. We deploy the two explainability methods as $E_{\phi} \in \{ E_{\phi_{1}}, E_{\phi_{2}} \}$ where $E_{\phi}$ accounts for the explainability methods parameterized by $\phi$ and $E_{\phi_{1}},E_{\phi_{2}}$ represent the GNNExplainer and PGExplainer, respectively. The two explainability methods are used to generate an explainbility edge mask $M^{v}_{E} \in \mathbb{R}^{|\mathcal{E}|}$ for each node $v \in \mathcal{V}$. The generated explanation mask $M^{v}_{E}$ can be defined with the following equation:
\begin{equation}
   M^{v}_{E} :=E_{\phi}( f_{\theta}, \mathcal{G},k,v)
\end{equation}
This equation shows that the explainability method takes in the input graph, trained model, the top-k value and a specific node and then generates the explanation edge mask for that node.
Note that, the generated explanation mask $M^{v}_{E}$ is binary in nature. Even though the initial generated mask is continuous, they are converted to a  binary mask through thresholding. Another component of the generated mask is sparsity which determines the number of edges identified as important. The sparsity is controlled by a top-k value hyperparameter. For a specific value of $k$, there will be $k$ entries in $M^{v}_{E}$ which are considered ``important" by the explainability method and are equal to 1. The rest of the entries in the explanation edge mask will be $0$ and considered unimportant. After the edge explanation mask for all nodes is generated a combined explanation mask $M_{E}$ for the overall graph is generated with \eqref{union}.
\begin{equation}
\label{union}
     M_{E} := \bigcup  \{ M^{v_1}_{E}, M^{v_2}_{E}, M^{v_3}_{E}, \cdots , M^{v_N}_{E}   \}
\end{equation}
The nodes connected to $M_{E}$ are treated as the important nodes generated by the explainability algorithm. For convenience, we represent these important nodes in a set denoted as $\mathcal{V}_{E}$.
This important node set $\mathcal{V}_{E}$ can be broken down into two overlapping sets $\mathcal{V}^{C_{ii}}_{E}$ and  $\mathcal{V}^{C_{ij}^{\prime}}_{E}$ where $\mathcal{V}^{C_{ii}}_{E}$ is the important node set that contains node from the same class (class $i$) and $\mathcal{V}^{C_{ij}^{\prime}}_{E}$ is the important node set that contains nodes from different classes (class $i$ and $j$). In our graph rewiring algorithm, the edge deletion is performed between the nodes in set $\mathcal{V}^{C_{ii}}_{E}$ and edge insertion is done between the nodes in $\mathcal{V}^{C_{ij}^{\prime}}_{E}$. From the set of edges between the nodes of $\mathcal{V}^{C_{ii}}_{E}$, a subset of edges is randomly selected according to a uniform distribution and it is followed by a deletion operation. For edge insertion, instead of edges, a subset of nodes is selected from the node set  $\mathcal{V}^{C_{ij}^{\prime}}_{E}$ according to a uniform distribution and edges are inserted between those nodes. Note that, our approach differs from just random edge insertion and edge deletion because we are performing edge insertion and deletion on a subset of $\mathcal{V}$ rather than the entire set of $\mathcal{V}$. 
\begin{algorithm}[H]
  \scriptsize
\caption{ \small Overall Algorithm for the Proposed Model}
\begin{algorithmic}[1]
\setstretch{1.5}
\Require  $\mathcal{G}$, $\mathcal{Y}$, $ f_{\theta}, E_{\phi}, \text{Hyperparameters of Model \& Explainability Algorithms} $ 
\Ensure Rewired Graph ($\mathcal{G}_r$)
\State  $\mathcal{Y}_{train},\mathcal{Y}_{test}, \mathcal{Y}_{val} \gets  \mathcal{Y} $
\State Initialize $\theta$ for $f_{\theta}$
 \vspace{1cm}
 \While{$ \text{epoch}< T_M $} 
\State $\mathbf{h}_v^{(L)} \gets f_{\theta}(\mathcal{G})$ 
\State $\text{Compute}~\mathcal{L}_{train}, \mathcal{L}_{val} \text{with}~\mathbf{h}_v^{(L)},\mathcal{Y}_{train},\mathcal{Y}_{val}$ 
\State $\text{Perform backpropagation and update model parameters}~\theta$
 \EndWhile
 \State $f_\theta \gets \text{Model with lowest}~\mathcal{L}_{val}$

 \For{$v = v_1$ to $v_{N}$}
 \State Initialize $\phi$ for $E_{\phi}$
 \State $\text{Train}~E_{\phi}~\text{for}~T_E ~\text{epochs}$
\State $M^{v}_{E} \gets E_{\phi}( f_{\theta}, \mathcal{G},k,v)$
\EndFor
\State  $ M_{E} \gets \bigcup  \{ M^{v_1}_{E}, M^{v_2}_{E}, M^{v_3}_{E}, \cdots , M^{ {v_N} }_{E}   \} $
\State $\mathcal{V}_E \gets \text{Get connected nodes to}~M_E$
\State $ \mathcal{V}^{C_{ii}}_{E},\mathcal{V}^{C_{ij}^{\prime}}_{E} \gets \mathcal{V}_E~\text{where}~i,j \in \{1,2,\cdots, m \}$
\State $\text{Random edge insertions between the nodes of } \mathcal{V}^{C_{ij}^{\prime}}_{E}   $
\State $\text{Random edge deletion between the nodes of }  \mathcal{V}^{C_{ii}}_{E}$
\State Return $\mathcal{G}_r$
\end{algorithmic}
\label{algo:training}
\end{algorithm}

\subsection{Attacker's Capability, Knowledge and Goal}\label{prev}
Background on the categorization of adversarial attack was briefly discussed in section~\ref{subsec:background_at}.

\begin{table}[!h]
\centering
\caption{Summary of Attacker's goal, capability and knowledge}
\renewcommand{\arraystretch}{1.1}
	\setlength{\tabcolsep}{9pt}
	\resizebox{\linewidth}{!}{
 \setlength{\extrarowheight}{.3em}
\begin{tabular}{c|c|c|c}
\hline
\textbf{\begin{tabular}[c]{@{}c@{}}Attacker's \\ Knowledge\end{tabular}} &
  \textbf{\begin{tabular}[c]{@{}c@{}}Targeted or\\  Untargeted\end{tabular}} &
  \textbf{\begin{tabular}[c]{@{}c@{}}Evasion or \\ Poisoning\end{tabular}} &
  \textbf{\begin{tabular}[c]{@{}c@{}}Perturbation\\ Type\end{tabular}} \\ \hline
White-Box &
  Untargeted &
  Evasion &
  \begin{tabular}[c]{@{}c@{}}Edge Deletion\\ and Insertion\end{tabular} \\ \hline
\begin{tabular}[c]{@{}c@{}}Have access\\ $f_{\theta}$ \\ which is required\\ for $E_{\phi}$\end{tabular} &
  \begin{tabular}[c]{@{}c@{}} Make use a single \\ global mask \\ $M_{E}$ to identify set of \\ nodes $\mathcal{V}_{E}$\end{tabular} &
  \begin{tabular}[c]{@{}c@{}}Edge insertion and \\ deletion is done\\ during test time, \\ not during training time\end{tabular} &
  \begin{tabular}[c]{@{}c@{}}Edge deletion done \\ between $\mathcal{V}^{C_{ii}}_{E}$ \& \\  edge insertion done \\ between $\mathcal{V}^{C^{\prime}_{ij}}_{E}$ \end{tabular} \\ \hline
\end{tabular}}
\label{tab:ackg}
\end{table}

Based on that, we can summarize the attacker's goal, capability and knowledge in Table~\ref{tab:ackg}. The attacker's goal is to perform edge perturbation \textit{i.e.} edge insertion and edge deletion. For the attacker to perform edge insertion and edge deletion between the nodes in $\mathcal{V}_{E}$, the attacker needs to know $M_{E}$ and for that attacker needs access to $f_{\theta}$ and $\mathcal{G}$ so that they can used in conjunction with a explainability method $E_{\phi}$. Having access to $f_{\theta}$ means a white-box attack. Another aspect is, that the attacker will try to decrease the test accuracy over all the test nodes, rather than a single node or a specific set of nodes. This implies an untargeted attack and also as modification is done during only test time it is an evasion attack.

\section{Numerical Tests}\label{sec:num_test}
In this section, we present the details about the datasets, training and numerical results followed by interpretation of the results.
\subsection{Dataset Description}
We consider three commonly node classification datasets\cite{sen2008collective}; Cora, CiteSeer and PubMed. All three are citation networks. The statistics for the number of classes, number of nodes and edges are summarized in Table~\ref{tab:dataset_stats}. In addition, the number of nodes in the training set, test set and validation set are also mentioned along with the number of intra-class and inter-class edges.

\begin{table}[htbp]
\centering
\caption{ Dataset Information and Statistics }
\renewcommand{\arraystretch}{0.9}
	\setlength{\tabcolsep}{9pt}
	\resizebox{1\linewidth}{!}{
 \setlength{\extrarowheight}{.3em}

\begin{tabular}{lccc}
\toprule
& \textbf{Cora} & \textbf{CiteSeer} & \textbf{PubMed}\\
\toprule
\textbf{Classification Settings} & Transductive & Transductive & Transductive\\
\textbf{Feature Dimension} &1433 & 3703 & 500\\
\textbf{\#~Classes} & 7 & 6 & 3\\ 
\vspace{3pt}
\underline{Node Information}: & & &\\
\textbf{\#~Nodes} & 2,708 & 3,327 & 19,717\\
\textbf{\#~Training Nodes} & 140 & 120 & 60\\
\textbf{\#~Test Nodes} & 1,000 & 1,000 & 1,000\\
\vspace{3pt}
\textbf{\#~Validation Nodes} & 500 & 500 & 500\\
\vspace{3pt}
\underline{Edge Information}: & & &\\
\textbf{\#~Edges} & 10,556 &  9,104 & 88,648\\
\textbf{\#~Intra-Class Edges} & 8,550 & 6,696 & 71,130 \\
\textbf{\#~Inter-Class Edges} &  2,006 &2,408 & 17,518\\

\bottomrule
\end{tabular}}
\label{tab:dataset_stats}
\end{table}

The edge numbers are reported considering a symmetric adjacency matrix and not considering self-loops. The number of intra-class edges and inter-class edges is important because of the nature of our proposed method and based on Table~\ref{tab:dataset_stats} for all three datasets, the number of intra-class edges is greater than inter-class edges which implies homophily. In the context of GNN, homophily refers to the tendency of nodes with similar attributes or features to be more connected than nodes with dissimilar attributes.

\subsection{Training Details and Hyperparameters}
In this section, the hyperparameters used for training the three models and additional model training-related information are introduced.
\begin{table}[!h]
\centering
\caption{Hyperparameters for Model Training and Optimization of Explainability Methods}
\renewcommand{\arraystretch}{1.2}
	\setlength{\tabcolsep}{9pt}
	\resizebox{\linewidth}{!}{
 \setlength{\extrarowheight}{.3em}

\begin{tabular}{lccc}
\toprule
     & GAT & GCN & GraphSAGE  \\ \midrule

\underline{\textbf{Model Hyperparameter:}}  &  &\\

 Number of Epochs~($T_M$)   & 2 & 2 & 2\\
 Number of Layers~($L$)   & 2 & 2 & 2\\

 \makecell[l]{Dimension of Hidden Layer Feature}~($d_{hidden}$) & 16 &  16 & 16\\ 

 Non-linear Activation Function~($\sigma$) & ELU  & ReLU  & ReLU \\ 

  Dropout  Rate                        & 0.6 & 0.5 & 0.5 \\
  Initial Learning Rate                & 0.01   & 0.01    & 0.01   \\ 
  Weight Decay                         & $5 \times 10^{-4}$   & $5 \times 10^{-4}$    & $5 \times 10^{-4}$   \\
  Number of Heads                      & 1   & -   & -  \\ \\

\toprule
     &  & GNNExplainer & PGExplainer  \\ \midrule
\underline{\textbf{Explainability Method Hyperparameter:}}  &  &\\
   Number of Epochs $(T_E)$               &  & 200 & 30 \\
     Learning Rate                &  & -   & 0.003 \\
    Thresholding Type               &  & top-k & top-k\\ 
\bottomrule
\end{tabular}}

\label{tab:Hyperparameter Table_table}
\end{table}
Each model was trained for $200$ epochs with an Adam\cite{kingma2014adam} optimizer with an initial learning rate of $0.01$. The hardware used for training was an NVIDIA A100 80GB GPU. To keep it consistent, the number of layers for each model is $2$ which provides a fair comparison between the models. Another rationale behind keeping the number of layers limited to $2$ is the over-smoothing issue~\cite{cai2020note} in deep GNNs. For deep GNNs, as the number of layers increases, the node embeddings start to become similar across all nodes. To minimize this, in practice, the number of layers is kept within the range of $2$ or $3$. All the hyperparameters for training the models and hyperparameters related to the explainability method are summarized in Table~\ref{tab:Hyperparameter Table_table}. To stabilize training and prevent issues with overfitting, L2 regularization is used with a weight decay parameter of $5\times 10^{-4}$. This makes sure the weights do not get too large as the training progresses. In addition, Dropout is used to prevent overfitting and increase the generalization ability of the models.
Note that, both the explainability method GNNExplainer and PGExplainer have their own internal optimization algorithm to generate the explanation mask. Hence, there is an epoch number $T_{E}$ and learning rate associated with both explainability algorithms. The code implementation utilizes \textit{PyTorch Geometric} a popular graph learning framework.

\begin{table*}[!t]
\centering
\caption{Results for our proposed Model with Reported Metrics as Misclassification Rate in (\%).  }
\renewcommand{\arraystretch}{1.2}
	\setlength{\tabcolsep}{9pt}
	\resizebox{0.8\linewidth}{!}{
 \setlength{\extrarowheight}{.3em}
\begin{tabular}{llccccccccccc}
\hline
\textbf{Dataset} &  & \multicolumn{3}{c}{\textbf{Cora}} & \multicolumn{1}{l}{} & \multicolumn{3}{c}{\textbf{CiteSeer}} & \multicolumn{1}{l}{} & \multicolumn{3}{c}{\textbf{PubMed}} \\ \cline{3-5} \cline{7-9} \cline{11-13} 
Models &  & \multicolumn{1}{l}{\textit{GAT}} & \multicolumn{1}{l}{\textit{GCN}} & \multicolumn{1}{l}{\textit{GraphSAGE}} & \multicolumn{1}{l}{} & \multicolumn{1}{l}{ \textit{GAT}} & \multicolumn{1}{l}{\textit{GCN}} & \multicolumn{1}{l}{\textit{GraphSAGE}} & \multicolumn{1}{l}{} & \multicolumn{1}{l}{\textit{GAT}} & \multicolumn{1}{l}{\textit{GCN}} & \multicolumn{1}{l}{\textit{GraphSAGE}} \\
Clean &  &  18.00\% & 18.80\%& 19.90\% &  & 30.40\%& 28.30\%& 30.90\% &  & 22.40\%& 20.70\% &22.70\%\\ \hline
\textbf{GNNExplainer} $(k=2)$ &  &  &  &  &  &  &  &  &  &  &  &  \\
$\gamma>1, EDR<5\% $ &  & 34.50\% &35.50\% & 34.30\% &  & 53.50\% & 51.70\% & 47.70\% &  & 38.80\%& 38.00\% & 44.80\% \\
$\gamma<1, EDR>>5\%$ &  & 28.80\% &  30.40\% &  33.30\% &  & 39.30\% & 39.60\%  & 41.90\% &  & 23.30\% & 23.50\% & 23.80\% \\
\hline
\textbf{PGExplainer} $(k=1000)$ &  &  &  &  &  &  &  & &  &  &  & \\
$\gamma>1, EDR < 5\%$ &  & 26.40\% & 27.10\%  & 24.90\% &  & 34.40\% & 38.80\% & 35.60\% &  & 24.10\% & 22.70\%& 23.10\% \\
$\gamma<1, EDR >> 5\%$&  & 25.90\% & 26.40\% & 25.20\% &  & 33.40\% & 36.50\% & 34.30\% &  &23.10\% & 21.60\% & 22.00\% \\
\hline
\end{tabular}}
\label{tab:results}
\end{table*}

\subsection{Tests Results}\label{sec:numerical_results}
 For numerical results, we used the misclassification rate as a metric.  The misclassification rate $MCR$ is defined as follows:
\begin{equation}
    MCR := 1-\frac{\sum \mathds{1}(\hat{\mathcal{Y}}_{test_{i}}=\mathcal{Y}_{test_{i}})}{{|\mathcal{V}_{test}|}}
\end{equation}
where, $\hat{\mathcal{Y}}_{test_{i}}$ is the predicted node label for the $i^{th}$ test node, $\mathcal{Y}_{test_{i}}$ is the ground truth node label on the test dataset for the $i^{th}$ node and $|\mathcal{V}_{test}|$ is the total number of test nodes. 

The goal is to increase the $MCR$ with the proposed graph rewiring method. Based on the statistics presented in Table \ref{tab:dataset_stats}, there are a lot more intra-class edges compared to inter-class edges. Therefore, the number of edge deletions required to have a large decrease in $MCR$ is large. In contrast, as the number of intra-class edges is small, with a few edge insertions between intra-class nodes significantly decreases the $MCR$. The results obtained are presented in Table~\ref{tab:results}. First, we report the $MCR$ in the ``clean" dataset without any perturbation in the graph.
This is essentially the baseline for comparison with the perturbed version of the graph.

\begin{figure}[!h]
    \centering
    \includegraphics[width=0.8\linewidth]{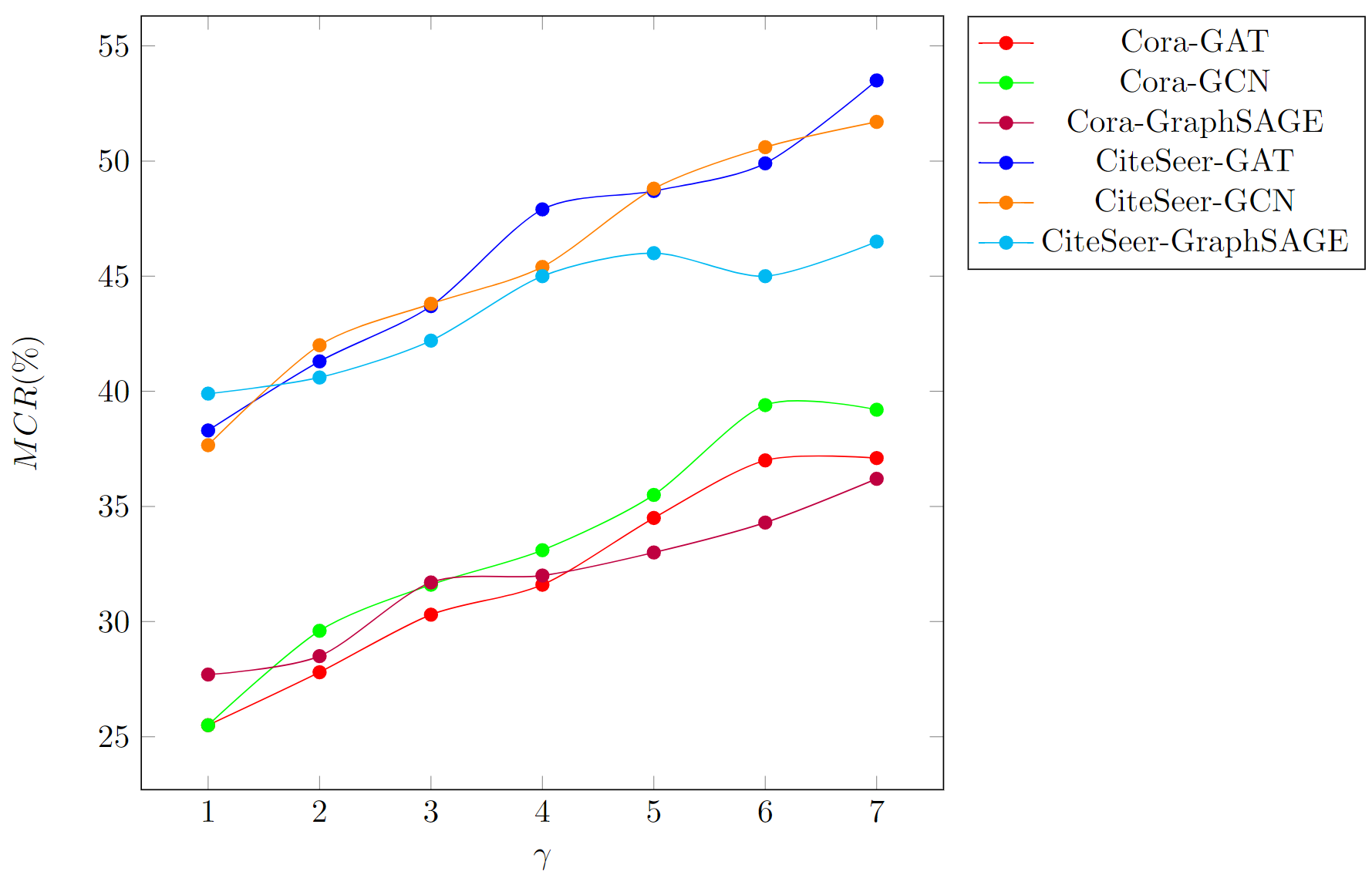}
    \caption{$MCR(\%)$ vs $\gamma$ plot with GNNExplainer as the explainability algorithm (For visual clarity, only Cora and CiteSeer are shown).}
    \label{fig:gamma}
\end{figure}

\begin{figure}[!h]
    \centering
    \includegraphics[width=0.8\linewidth]{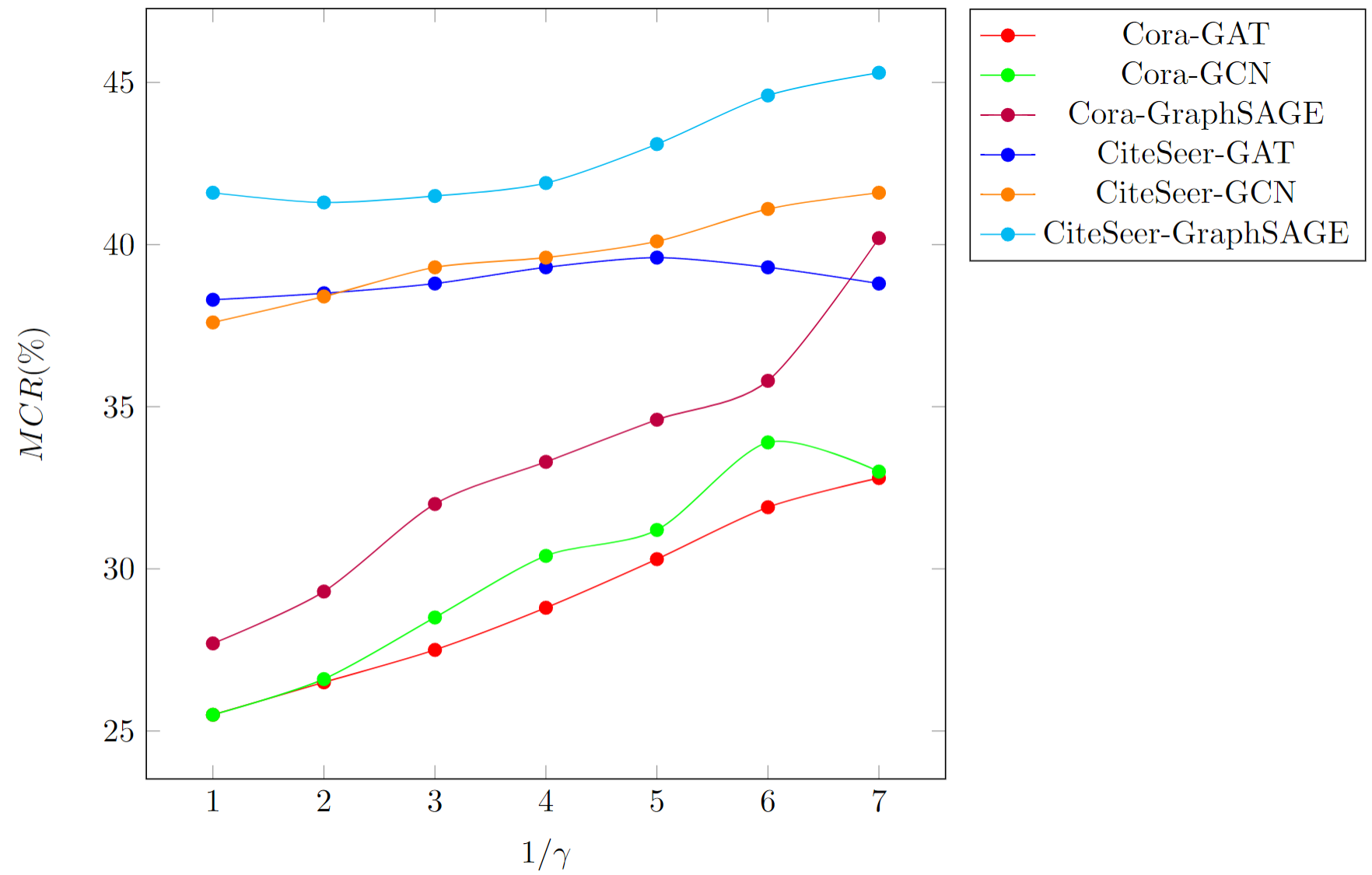}
    \caption{$MCR(\%)$ vs $1/\gamma$ plot with GNNExplainer as the explainability algorithm (For visual clarity, only Cora and CiteSeer are shown).}
    \label{fig:one_by_gamma}
\end{figure}

The three hyperparameters varied to generate the result are the top-k value of the explainability method, the ratio of edge insertions and deletions denoted by $\gamma$ and $EDR$ which stands for edge difference rate. The $\gamma$ parameter is defined as below:

\begin{equation}
    \gamma := \frac{\text{\# edge insertions between the nodes of}~\mathcal{V}^{C_{ij}^{\prime}}_{E}}{\text{\# edge deletion between the nodes of}~\mathcal{V}^{C_{ii}}_{E}}
\end{equation}

The $\gamma$ parameter is a good indicator of how effective edge insertion is compared to edge deletion.

\begin{figure*}
    \centering
    \includegraphics[width=0.9\linewidth]{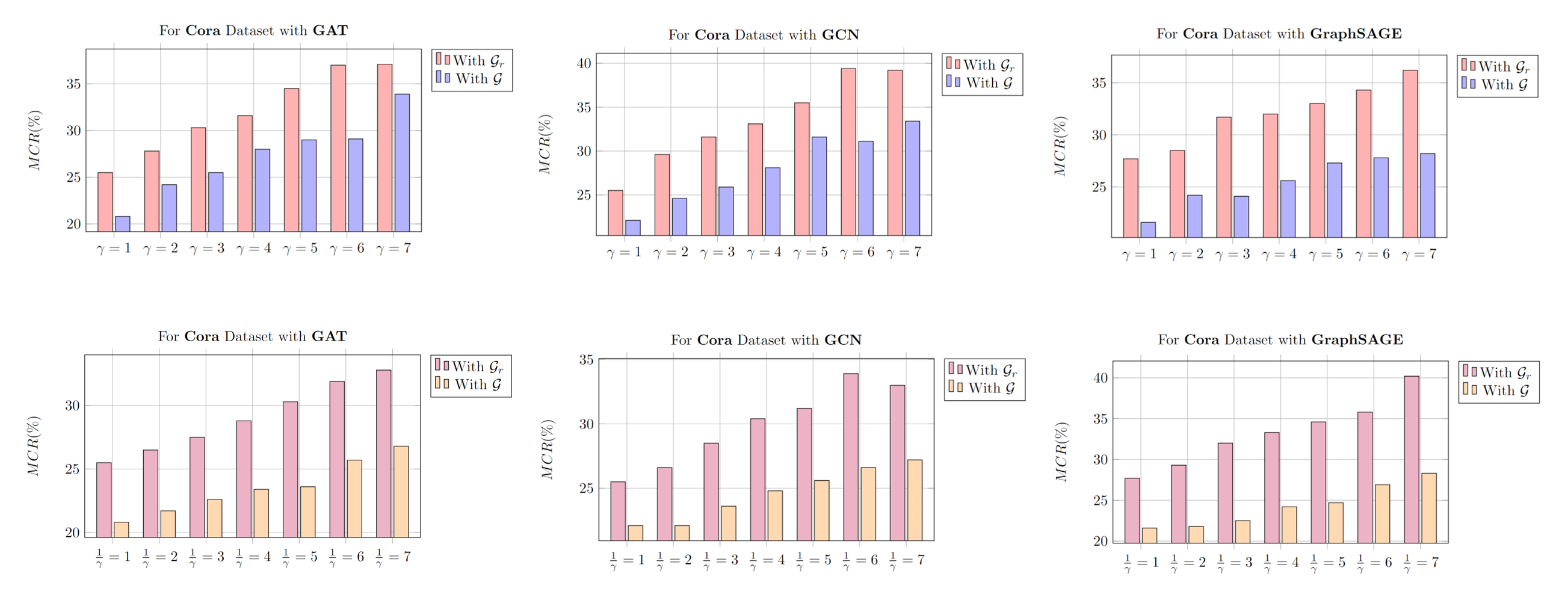}
    \caption{(\textbf{Top row}) $MCR$ vs $\gamma$ plot with Cora dataset showing comparison between original Graph $\mathcal{G}$ and rewired graph $\mathcal{G}_r$ with GNNExplainer. (\textbf{Bottom row}) $MCR$ vs $1/\gamma$ plot with Cora dataset showing comparison between original Graph $\mathcal{G}$ and rewired graph $\mathcal{G}_r$ with GNNExplainer.}
 \label{fig:ex_gamma}
\end{figure*}
\begin{figure*}[!h]
    \centering
    \includegraphics[width=0.7\linewidth]{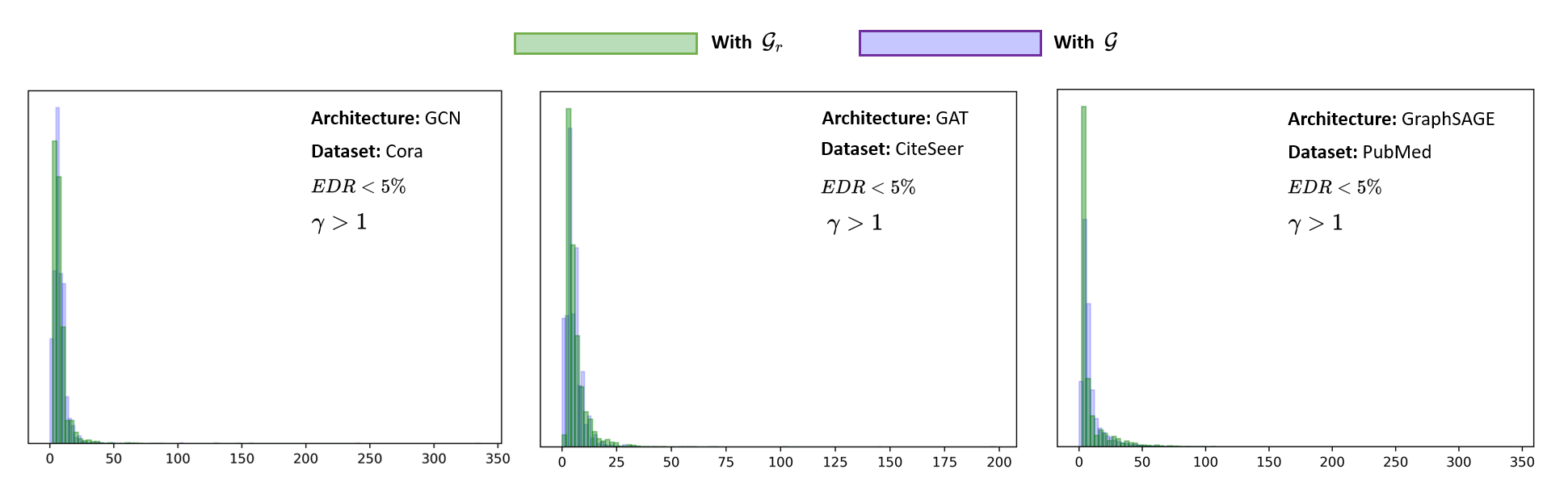}
    \caption{Node degree Distribution for the three datasets with highest $MCR$ value. For all three instances $\gamma>1$, $EDR<5\%$ and the explainability method is GNNExplainer. (\textbf{Left}) The dataset is Cora and the architecture is GCN. (\textbf{Middle}) The dataset is CiteSeer and the architecture is GAT. (\textbf{Right}) The dataset is PubMed and the architecture is GraphSAGE.}
    \label{fig:degree_distribution}
\end{figure*}

A value of $\gamma>1$ indicate the number of edge insertion between the nodes of $\mathcal{V}^{C_{ij}^{\prime}}_{E}$ is greater than the number of edge deletion between the nodes of $\mathcal{V}^{C_{ii}}_{E}$. In contrast,  $\gamma<1$ indicate the number of edge deletion between the nodes of $\mathcal{V}^{C_{ii}}_{E}$ is greater than the number of edge insertion between the nodes of $\mathcal{V}^{C_{ij}^{\prime}}_{E}$.

On the other hand, $\gamma=1$ signifies the number of edge insertions is equal to the number of edge deletions. Another parameter $EDR$ is the edge count difference between the original graph $\mathcal{G}$ and the rewired version $\mathcal{G}_r$. Due to the nature of our proposed algorithm, the value of $EDR$ is not fixed across different datasets and different classifiers $f_{\theta}$ as the size of the node set $\mathcal{V}_{E}$ \textit{i.e.} $\lvert \mathcal{V}_{E} \rvert$ varies. Hence, to provide a fair and reasonable comparison,  we report $EDR$ where it is less than and more than $5\%$ across all the datasets and classifiers. Note that, as no feature perturbation is done, $\mathbf{X}=\mathbf{X}_{r}$ \textit{i.e.} the feature matrix of the original graph  $\mathcal{G}$ and the rewired graph $\mathcal{G}_r$ remains the same. From Table~\ref{tab:results}, comparing the values between $\gamma>1$ and $\gamma<1$, it is apparent that, the $MCR$ is higher for $\gamma>1$ compared to $\gamma<1$ which verifies our assumptions about the effectiveness of edge insertions against edge deletions. Also, with edge insertions, the $EDR$ can be kept lower compared to edge deletions.

Comparing the $MCR$ across GAT, GCN and GraphSAGE, one important conclusion we can make is that GraphSAGE is much more effective for large graphs like PubMed. For comparatively small graphs like Cora and CiteSeer GCN and GAT is much more effective.

If we make a comparison between the explainability algorithms GNNExplainer and PGExplainer, it is evident that GNNExplainer has a greater impact in increasing the $MCR$ compared to PGExplainer. Even with a top-k value of $1000$, the $MCR$ for PGExplainer is lower compared to GNNExplainer. This can be attributed to the fact that PGExplainer is optimized for inductive training as opposed to transductive training.

To further validate the effectiveness of edge insertions compared to edge deletions we ran further analysis by varying the value of $\gamma$ and the results are presented in Fig.~\ref{fig:gamma} and Fig.~\ref{fig:one_by_gamma}. In Fig.~\ref{fig:gamma}, the $MCR$ value is plotted against the increasing value of $\gamma$ for the Cora and CiteSeer datasets. All three classifiers show an upward trend will slight fluctuations. The slight random fluctuations arise owing to the random edge insertions and deletions. Also, due to the baseline $MCR$ on the clean data being high for the CiteSeer dataset, it also has a higher $MCR$ value compared to the Cora dataset.

For Fig.~\ref{fig:one_by_gamma} the $MCR$ is plotted against $1/\gamma$ instead $\gamma$ to provide comparison with Fig.~\ref{fig:gamma}. A higher value of $1/\gamma$ signifies number of edge deletions is greater than edge insertions. It also shows an upward trend with an increase in $1/\gamma$ but the rate of increase is much less and hence flatter compared to the curves in Fig.~\ref{fig:gamma}.
\begin{table*}[t]
\caption{Comparison of Results with Existing Works}
\centering
\renewcommand{\arraystretch}{1.3}
	\setlength{\tabcolsep}{9pt}
	\resizebox{0.9\linewidth}{!}{
 \setlength{\extrarowheight}{.3em}
\begin{tabular}{|lccc|c|cc|cc|cc|}
\hline
\multicolumn{1}{|c}{\multirow{2}{*}{\textbf{}}} &
  \multirow{2}{*}{\textbf{\begin{tabular}[c]{@{}c@{}}Xu \etal~\cite{xu2019topology}\\ (5\% perturbation)\end{tabular}}} &
  \multirow{2}{*}{\textbf{\begin{tabular}[c]{@{}c@{}}Z{\"u}gner \etal~\cite{zugner2019adversarial}\\ (5\% perturbation)\end{tabular}}} &
  \multirow{2}{*}{\textbf{Ours}} &
  \multirow{2}{*}{\textbf{Dataset}} &
  \multicolumn{2}{c|}{\textbf{Xu \etal~\cite{xu2019topology}}} &
  \multicolumn{2}{c|}{\textbf{Z{\"u}gner \etal~\cite{zugner2019adversarial}}} &
  \multicolumn{2}{c|}{\textbf{Ours}} \\ \cline{6-11} 
\multicolumn{1}{|c}{} &
   &
   &
   &
   &
  Cora &
  CiteSeer &
  \multicolumn{1}{c}{Cora} &
  \multicolumn{1}{c|}{CiteSeer} &
  \multicolumn{1}{c}{Cora} &
  \multicolumn{1}{c|}{CiteSeer} \\ \hline
Targeted &
  \ding{55} &
  \ding{55} &
  \ding{55} &
  \multirow{2}{*}{Architecture} &
  \multirow{2}{*}{GCN} &
  \multirow{2}{*}{GCN} &
  \multirow{2}{*}{DeepWalk} &
  \multirow{2}{*}{DeepWalk} &
  \multirow{2}{*}{GCN} &
  \multirow{2}{*}{GAT} \\ \cline{1-4}
Untargeted &
  \ding{51} &
  \ding{51} &
  \ding{51} &
   &
   &
   &
   &
   &
   &
   \\ \hline
Black-Box &
  \ding{55} &
  \ding{55} &
  \ding{55} &
  \multirow{2}{*}{Approach} &
  \multirow{2}{*}{CE-min-max} &
  \multirow{2}{*}{CW-min-max} &
  \multirow{2}{*}{Meta-Self} &
  \multirow{2}{*}{Meta-Self} &
  \multirow{2}{*}{\makecell[c]{Proposed \\ Method}} &
  \multirow{2}{*}{\makecell[c]{Proposed \\ Method}} \\ \cline{1-4}
White-Box &
  \ding{51} &
  \ding{55} &
  \ding{51} &
   &
   &
   &
   &
   &
   &
   \\ \hline
Gray Box &
  \ding{55} &
  \ding{51} &
  \ding{55} &
  \multirow{3}{*}{$MCR$} &
  \multirow{3}{*}{$31.00\%$} &
  \multirow{3}{*}{$40.00\%$} &
  \multirow{3}{*}{$32.70\%$} &
  \multirow{3}{*}{$46.30\%$} &
  \multirow{3}{*}{$^{\dagger}\mathbf{33.10\%}$} &
  \multirow{3}{*}{$^{\dagger}\mathbf{47.80\%}$} \\ \cline{1-4}
Evasion &
  \ding{51} &
  \ding{55} &
  \ding{51} &
   &
   &
   &
   &
   &
   &
   \\ \cline{1-4}
Poisoning &
  \ding{51} &
  \ding{51} &
  \ding{55} &
   &
   &
   &
   &
   &
   &
   \\ \hline 
\multicolumn{11}{l}{$^{\dagger} MCR~\text{adjusted to account for baseline difference with  \cite{xu2019topology} and \cite{zugner2019adversarial}}$}
\end{tabular}}
\label{tab:comparison}
\end{table*}
\begin{figure*}
    \centering
    \includegraphics[width=0.7\linewidth]{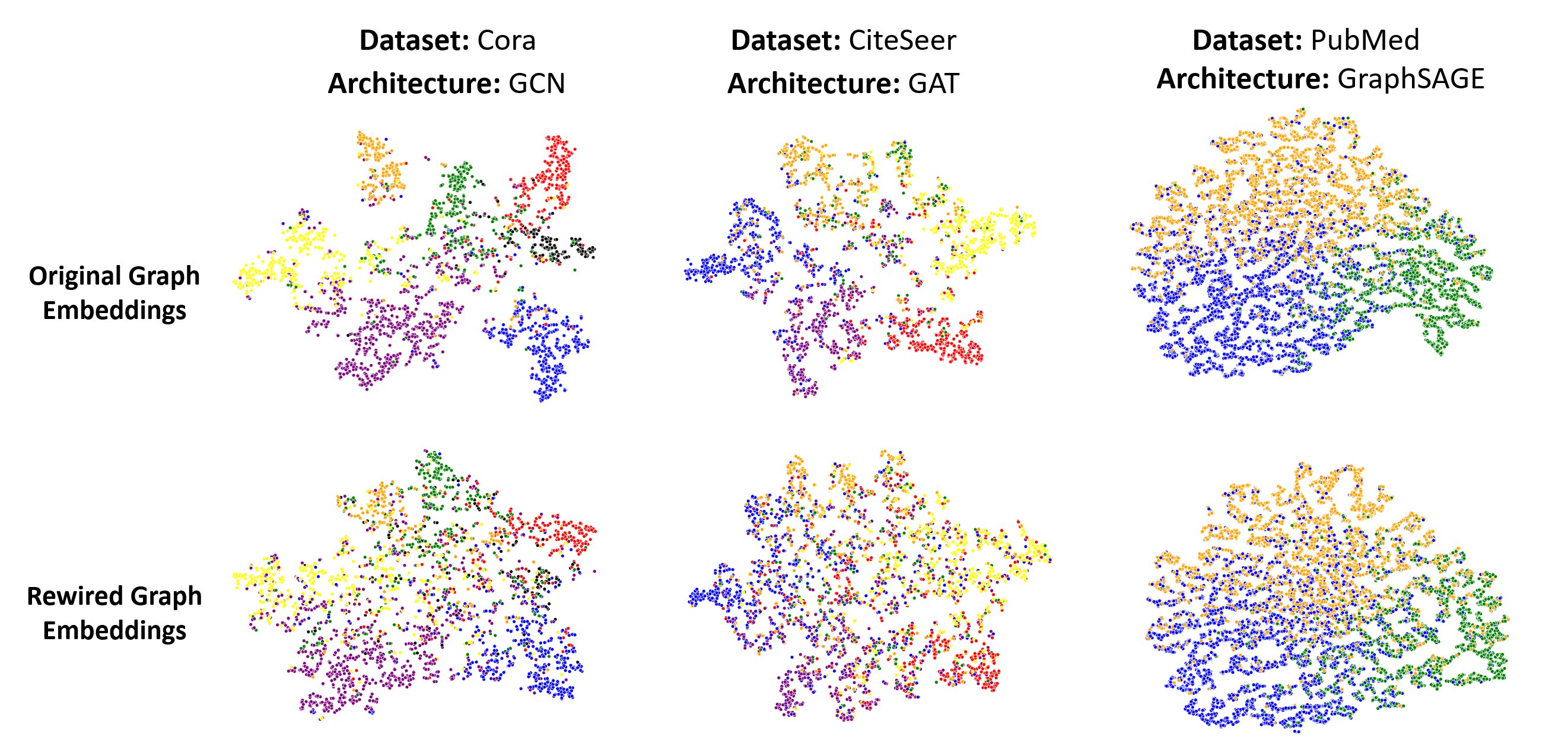}
    \caption{t-Distributed Stochastic Neighbor Embedding (t-SNE) plot of node embeddings for the original graph $\mathcal{G}$ (\textbf{Top Row}) and the rewired graph $\mathcal{G}_r$ (\textbf{Bottom Row}). For all three instances $\gamma>1$, $EDR<5\%$ and the explainability method is GNNExplainer.}
    \label{fig:t_sne}
\end{figure*}
To validate the use of explainability, we also compare the $MCR$ with and without using explainability methods. The results from this analysis are shown in Fig.~\ref{fig:ex_gamma}. To get the $MCR$ value for the without explainability method we remove the process of identifying important node set $\mathcal{V}_E$ and perform the edge insertions and deletion operation on the original graph node set $\mathcal{V}$ directly. For the sake of conciseness, results are reported with GNNExplainer as the explainability method. First, we perform this analysis with incremental $\gamma$ value where it starts at $\gamma=1$ and ends at $\gamma=7$. From Fig.~\ref{fig:ex_gamma} we can see that, for all values of $\gamma$, the $MCR$ is higher when edge perturbations are done between the nodes of $\mathcal{V}_E$. This unequivocally proves that identifying the important set $\mathcal{V}_E$ does indeed have an effect.
This observation holds for incremental $1/\gamma$ value also, but $MCR$ values are smaller compared to the previous analysis which further consolidates the effectiveness of edge insertions against edge deletions.  

The unnoticeability constraint for adversarial attack for the graph domain is different from the other domains like image. For the graph domain, the node degree distribution needs to be similar to the original graph node degree distribution. In Fig.~\ref{fig:degree_distribution} we plot the node degree distribution for all three datasets with the highest $MCR$ value reported in Table~\ref{tab:results}. There is no apparent noticeable difference between the original graph $\mathcal{G}$ and rewired graph $\mathcal{G}_r$.

We also compare our work with existing works in Table~\ref{tab:comparison}. Both Xu~\etal~\cite{xu2019topology} and Z{\"u}gner~\etal~\cite{zugner2019adversarial} perform only edge perturbation for node classification tasks. 

Xu~\etal used a gradient-based PGD attack with four different variants. Negative cross-entropy loss via PGD attack (CE-PGD), CW loss via PGD attack (CW-PGD), negative cross-entropy loss via min-max attack (CE-min-max), CW loss via min-max attack (CW-min-max). The authors only use GCN as the architecture. We compare with CE-min-max for the Cora dataset and CW-min-max for the CiteSeer. 

Z{\"u}gner~\etal uses meta-gradients for the adversarial attack with three different architectures GCN, DeepWalk~\cite{perozzi2014deepwalk} and Column Network~\cite{pham2017column}. We provide a comparison with the DeepWalk architecture used in their work where one of their approach is Meta-Self.

To establish a fair comparison, we first compare the attacker's capability, knowledge and goal. It's important to note that both Xu~\etal and Z{\"u}gner~\etal perform an untargeted attack and require some knowledge of the model parameters meaning not a black-box attack.
  Also, we make sure to account for the baseline difference with~\cite{xu2019topology} and~\cite{zugner2019adversarial} by adjusting the $MCR$ value. To adjust for that we subtract the baseline difference from the $MCR$ value. It's worth highlighting that even though Xu~\etal uses their approach for both evasion and poisoning type attacks, Z{\"u}gner~\etal only does it for poisoning type attacks.

To corroborate our idea about feature contamination with the graph rewiring algorithm, we perform a dimensionality reduction technique t-SNE on the node embeddings and the generated plots are shown in Fig.~\ref{fig:t_sne}. The generated plots align with our claim because the embeddings for the perturbed version of the graph have more overlap compared to the unperturbed version.
\section{Conclusion} \label{sec:conclusion}
In this paper, we deal with the effect of edge perturbations on GNNs as an untargeted, white-box, evasion attack. Our approach involved identifying the key nodes in the graph through two different explainability methods and performing edge perturbations between the identified key nodes. Numerical results confrim the impact of the proposed approach. Furthermore, the effectiveness of two different types of perturbation as edge deletion and edge insertions were evaluated. The analysis and numerical tests indicate that edge insertions among nodes from different classes are more effective in increasing the $MCR$ as opposed to edge deletion between the nodes of the same class.

\bibliography{ref.bib}
\bibliographystyle{IEEEtran}

\end{document}